\documentclass[10pt, conference, compsocconf]{IEEEtran}
\IEEEoverridecommandlockouts
\usepackage{cite}
\usepackage[pdftex]{graphicx}
\usepackage{amsmath,amssymb,amsfonts}
\usepackage{algorithmic}
\usepackage{multirow}
\usepackage[normalem]{ulem}
\useunder{\uline}{\ul}{}
\usepackage[T1]{fontenc}
\usepackage{subfigure}
\usepackage{diagbox}
\usepackage{url}
\usepackage{booktabs}

\newtheorem{myDef}{Definition}

\begin{document}
%
\title{Hyper Meta-Path Contrastive Learning for Multi-Behavior Recommendation}

%
\author{\IEEEauthorblockN{Haoran Yang\IEEEauthorrefmark{1}\thanks{\IEEEauthorrefmark{1}: Equal contribution}\IEEEauthorrefmark{5},
Hongxu Chen\IEEEauthorrefmark{1}\IEEEauthorrefmark{2}\thanks{\IEEEauthorrefmark{2}: Corresponding author}\IEEEauthorrefmark{5},
Lin Li\IEEEauthorrefmark{3}, 
Philip S. Yu\IEEEauthorrefmark{4} and
Guandong Xu\IEEEauthorrefmark{2}\IEEEauthorrefmark{5}}
\IEEEauthorblockA{\IEEEauthorrefmark{5}School of Computer Science, University of Technology Sydney, Sydney, Australia}
\IEEEauthorblockA{\IEEEauthorrefmark{3}School of Computer and Artificial Intelligence, Wuhan University of Technology, Wuhan, China}
\IEEEauthorblockA{\IEEEauthorrefmark{4}Department of Computer Science, University of Illinois at Chicago, Chicago, U.S.A}
\IEEEauthorblockA{haoran.yang-2@student.uts.edu.au, \{hongxu.chen, guandong.xu\}@uts.edu.au \\
cathylilin@whut.edu.cn, psyu@cs.uic.edu}
}


\maketitle

\begin{abstract}
User purchasing prediction with multi-behavior information remains a challenging problem for current recommendation systems. Various methods have been proposed to address it via leveraging the advantages of graph neural networks (GNNs) or multi-task learning. However, most existing works do not take the complex dependencies among different behaviors of users into consideration. They utilize simple and fixed schemes, like neighborhood information aggregation or mathematical calculation of vectors, to fuse the embeddings of different user behaviors to obtain a unified embedding to represent a user's behavioral patterns which will be used in downstream recommendation tasks. To tackle the challenge, in this paper, we first propose the concept of hyper meta-path to construct hyper meta-paths or hyper meta-graphs to explicitly illustrate the dependencies among different behaviors of a user. How to obtain a unified embedding for a user from hyper meta-paths and avoid the previously mentioned limitations simultaneously is critical. Thanks to the recent success of graph contrastive learning, we leverage it to learn embeddings of user behavior patterns adaptively instead of assigning a fixed scheme to understand the dependencies among different behaviors. A new graph contrastive learning based framework is proposed by coupling with hyper meta-paths, namely HMG-CR, which consistently and significantly outperforms all baselines in extensive comparison experiments.

\end{abstract}


%
\IEEEpeerreviewmaketitle

\section{Introduction}
\label{sec:introduction}
Online shopping is becoming more and more essential nowadays, which generates a large volume of user behavioral data depicting users' purchasing motivations, interests, behavioral patterns, etc. However, many traditional recommendation systems \cite{ncf, graphsage} pay significant attention to purchasing alone, leaving other associated behavioral data unexploited. Though recent works \cite{meta-graph-rec, rgcn, nmtr, ehcf, sptf} reveal the gap and try to leverage multi-behavior information to improve recommendation quality, there are still limitations. For instance, some path-based works \cite{meta-graph-rec, temp-meta-path} leverage meta-paths \cite{metapath, meta-graph} to extract recommendation context to better characterise users' multiple behaviors. However, there exist many meta-path schemes observed in heterogeneous graphs, resulting in the difficulty of finding out the best one from multiple meta-path schemes via exhaustive search  or learning a specific rule or scheme from the heterogeneous graphs to construct meta-paths \cite{temp-meta-path}. Selecting effective and meaningful meta-path scheme in this case is time-consuming, and the reasons of the selection are usually unknown. 
\begin{figure}[htbp]
	\centering
	\subfigure[Meta-paths of a user]{
		\includegraphics[width=0.2\textwidth]{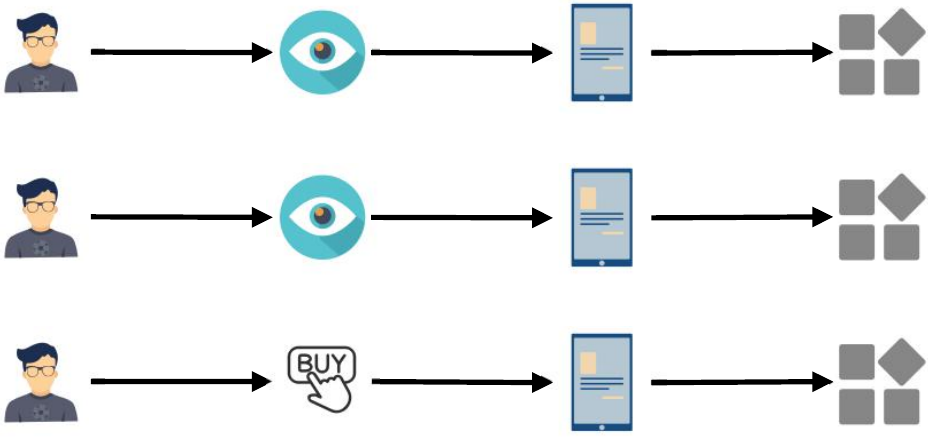}
	}
	\quad
	\subfigure[A hyper meta-path of a user]{
		\includegraphics[width=0.2\textwidth]{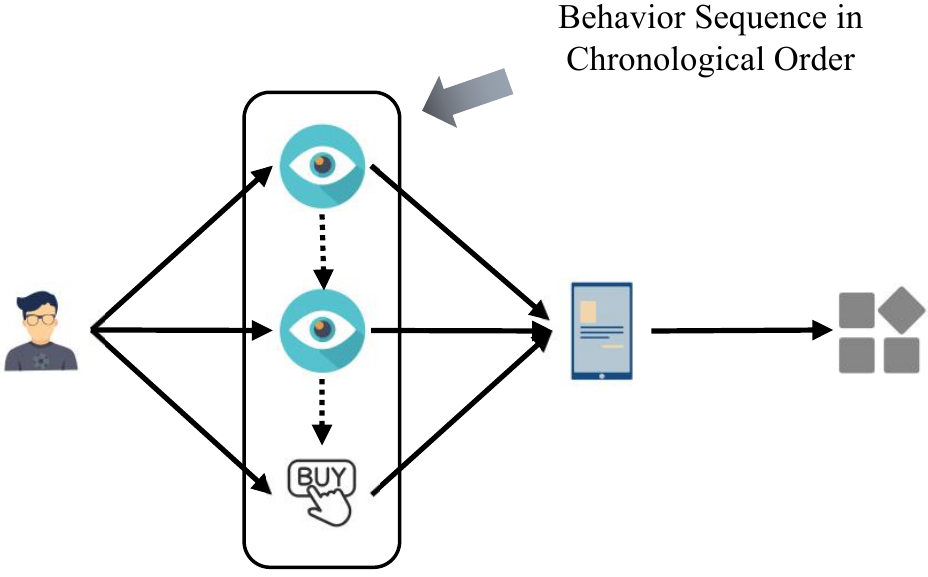}
	}
	\caption{(a) Meta-paths of a user in the recommendation system denotes the user's different behaviors. (b) A hyper meta-path constructed by previous meta-paths denotes the user's behavior pattern when he is purchasing a smartphone.}
	\label{hyper-meta-path}
\end{figure}

To overcome the above limitation of existing path-based approaches, we propose a new concept of hyper meta-path that consolidates multiple paths in a well-organised and holistic way. Similar to hyperedge in hypergraph \cite{hete-hyper-emb, multi-level-hyper, hyper-sr, inter-hypergraph} where an edge can connect more than two nodes, a hyper meta-path is a composition of multiple meta-paths between specified two end nodes in a heterogeneous network. As shown in Figure \ref{hyper-meta-path}, let us assume that before purchasing a phone, the user had viewed the item twice. If meta-path based approaches are adopted to model the different shopping-related behaviors, three independent meta-path instances (Fig. \ref{hyper-meta-path}.(a)) can be discovered to characterise this purchasing context. Though from these three meta-paths we can learn the user has viewed and purchased the item, it dose not explicitly  reflect the exact behavioral pattern of the user. That is, it is unclear whether the user purchased the item directly or viewed the item carefully before purchasing. 

In contrast, the proposed hyper meta-path is capable of achieving such a goal. As shown in part (b) of Fig. \ref{hyper-meta-path}, a hyper meta-path between the user and the phone consolidates all three behaviors in a sequential order, which explicitly shows that the user carefully viewed the phone twice before the final purchasing action. Note that the way of consolidating related meta-paths into a hyper meta-path can be flexible and generalized to any reasonable rule depending on particular application scenario. Besides, the concept of hyper meta-path is also useful for differentiating different behavior patterns between different users or when facing various categories of items presented to a user. For instance, normally technical people may research on different substitutable electronic products and take longer time to compare them, while non-technical people are not keen on investigating them and would probably directly buy one based on someone's recommendation. Even for the same user no matter their age and gender, they usually exhibit quite diverged buying patterns when facing different categories of products. For example, a user may have totally different buying patterns when purchasing large items (e.g., white goods like fridge or TV) and small fast-moving consumers goods (e.g., periodically buying tissues from online market without viewing again and again). 

Nevertheless, it is not straightforward to incorporate the modelling of hyper meta-path into existing learning frameworks. Currently, graph-based unsupervised learning approaches are mainly used for path-based recommendation. For example, GNNs based approaches \cite{rgcn} are a popular means for multi-behavior recommendation via aggregating information passed from different types of edges or nodes in heterogeneous information networks \cite{hin-survey}. Despite its popularity, these methods usually fuse the learned features of different behaviors independently, which is too naive to reflect hyper meta-path context for recommendation. Moreover, multi-task learning based models \cite{nmtr, ehcf} are also possible ways that introduce additional supervision signals from the observed multiple behavior data to improve recommendation quality. However, extra efforts on well-elaborated tasks are tricky, and researchers have to carefully work out the effective dependencies among related tasks. For example, taking purchasing prediction as a primary task while modelling the \textit{add-to-cart} behavior prediction as an auxiliary task might not be always right as some users may buy some items directly without putting them into the cart. 


Thus, to further reveal and capture the differences between buying patterns, together with hyper meta-paths, we innovatively leverage graph contrastive learning \cite{gcc, graph-aug} paradigm for the multi-behavior recommendation problem.
The main idea of graph contrastive learning is to distinguish the differences among graphs to obtain the useful structure information of each graph, raising a recent surge of interest \cite{graph-aug, gcc}.
The rationale of incorporating contrasitive learning with our proposed hyper meta-paths is that, a user may have multiple hyper meta-paths explicitly illustrating his/her behavioral patterns when facing different products. Since hyper meta-path explicitly describes users' behaviors towards purchasing different items, graph contrastive learning becomes a best fit for comparing and extracting the key structures in the graph consisting of hyper meta-paths. 



More specifically, we combine multiple hyper meta-paths of a user to construct several hyper meta-graphs. Each hyper meta-graph contains different number of types of behaviors, For example, the first hyper meta-graph contains \textit{buy} and the second hyper meta-graph contains \textit{buy} and \textit{page view}. In this case, different hyper meta-graphs reflect different behavioral patterns regarding different products of the user. Then, we conduct graph contrastive learning among the constructed hyper meta-graphs to adaptively obtain the complex dependencies among different behaviors and the embeddings representing different behavioral patterns.
For instance, in HMG-CR, we first build the target contrastive graph that only contains \textit{buy} interactions between users and items as it is the target behavior for recommendation systems, and the other contrastive hyper meta-graphs are added for comparison by incrementally introducing auxiliary behaviors to the precedent hyper meta-graph. After that, we conduct graph contrastive learning between the constructed contrastive hyper meta-graphs to successively obtain progressive and comprehensive representations for each types of behaviors. Finally, the recommendation will be performed based on those discovered behavior patterns and features.

The contributions of this paper can be summarized into three aspects:
\begin{itemize}
	\item We propose the concept of hyper meta-path explicitly illustrating the logical relations among a collection of meta-paths, which tackles the limitation of meta-path that is insufficient to model the interactions among meta-paths. Hyper meta-path can be regarded as an approach to enrich graph structures.
	\item We innovatively utilize graph contrastive learning to capture the complex behavior patterns of users adaptively, alleviating existing methods' limitation.
	\item We propose a novel and flexible recommendation framework coupling graph contrastive learning with hyper meta-path, which achieves the superior performances in the comprehensive comparison experiments.
\end{itemize}

\section{Preliminaries}
\label{sec:preliminaries}
This section will introduce some preliminaries regarding the  proposed concept of hyper meta-path.

\subsection{Meta-Path}
Heterogeneous networks have been intensively studied by a lot of researchers due to its ability of utilizing multi-model multi-typed graph data. To illustrate the power of hetergeneous networks, Sun et al. \cite{metapath} proposed the concept of meta-path, which is widely used by many existing works \cite{metapath2vec, meta-graph}  in the research area of heterogeneous networks modeling.
Each meta-path captures the features among the nodes on the meta-path from a particular semantic perspective. Due to the diversity of meta-paths in a heterogeneous graph, for the target (e.g., a node or an edge), there exist multiple meta-paths. Thus, the informative meta-paths give heterogeneous network models the chance to obtain the multi-model multi-typed features of nodes and their relations. This kind of data structure indeed shows the advantage in many real-world graph data mining applications \cite{semi-metagraph, hetesmi}. However, there are limitations existing in meta-paths mentioned in Section \ref{sec:introduction}, failing to capture the interaction information among multiple meta-paths.

\subsection{Hyper Meta-Path}
\label{sec:hmp}
Though we can build extra meta-paths based on the interactions among existing meta-paths, we cannot take an exhaustive method to compute on every meta-paths since the computation complexity is unaffordable. Inspired by the concepts of hyperedge and hypergraph, we find a way to integrate interaction information among meta-paths into the target.
According to the limitation of conventional meta-path mentioned above and advantages of hyperedge and hypergraph, we propose the concept of hyper meta-path to capture meta-path features and interaction information among them simultaneously.

\begin{myDef}
	\textbf{Hyper meta-path}. A \textit{hyper meta-path} is a logical composition of multiple meta-path schemas connecting two end nodes in a heterogenous information network. Hyper meta-path has the following properties:
	\begin{itemize}
		\item It describes the logical relations (e.g., chronological order, spatial order and topological order) among a sort of meta-paths with the same end nodes.
		\item Multiple hyper meta-paths, which have the same start node, compose a \textit{hyper meta-graph}. 
	\end{itemize}
\end{myDef}


\section{Methodology}
\begin{figure*}[htbp]
	\centering
	\includegraphics[width=0.8\textwidth]{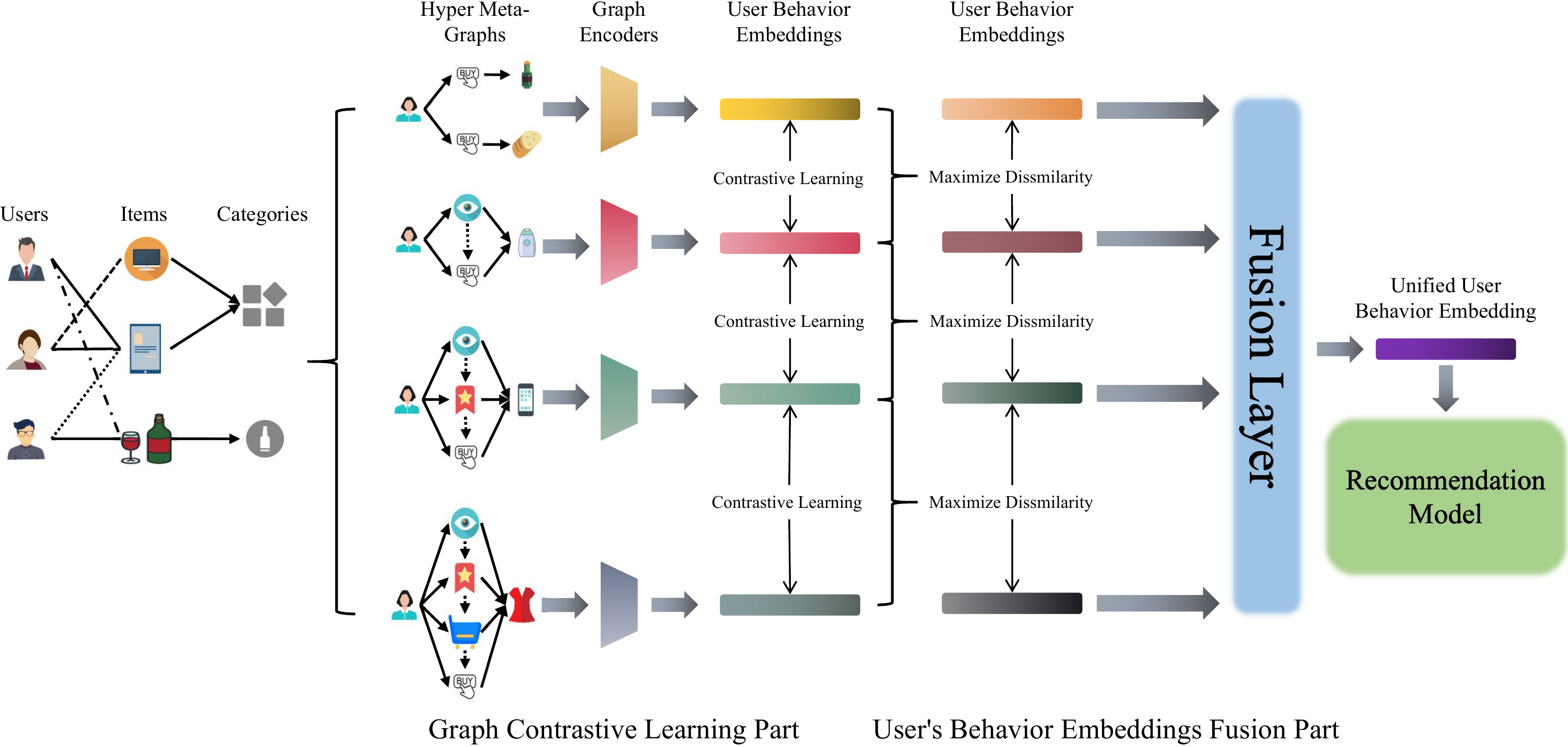}
	\caption{The overview of hyper meta-graph contrastive learning framework for recommedantions.}
	\label{hmg-cr}
\end{figure*}

In this section, we will introduce details of our proposed framework as shown in Figure \ref{hmg-cr}, \textbf{H}yper \textbf{M}eta-\textbf{G}raph \textbf{C}ontrastive learning for \textbf{R}ecommendations, namely HMG-CR.

\subsection{Hyper Meta-Graph Generation}
\label{sec:generation}
GNNs based recommendation methods \cite{lightgcn} have recently achieved tremendous success due to the power of GNNs. Data is critical to neural network models' performances. One way to leverage GNN based recommendation models is to construct proper graphs for them. The most common way to construct graphs in recommendation systems is building bipartite graphs via user-item interaction history. Since user-item interaction graphs are bipartite graphs, they are lack of semantic information because of their simple structure. To tackle this limitation, researchers have taken measures to further enrich semantic information carried by graphs. For example, adding side information into the graph \cite{side}, utilizing meta-paths existing in the graph \cite{temp-meta-path}, and constructing more complex graph structures (e.g., hypergraph) \cite{seq-hypergraph, unified-hypergraph}.

To improve recommendation results, in our work, we utilize the proposed concept of hyper meta-path to construct hyper meta-graphs carrying rich semantic information. Next we will introduce details to construct hyper meta-graphs for our proposed recommendation framework.

Given a set of interaction records in a recommendation system, $\{(u_j, r_k, i_q)|u_j\in\mathbf{U}, i_q\in\mathbf{I}, r_k\in\mathbf{R}\}$, where $\mathbf{U}=\{u_0, u_1, \cdots, u_n\}$ denotes the set of all users, $\mathbf{I}=\{i_0, i_1, \cdots, i_m\}$ denotes the set of all items, and $\mathbf{R}=\{r_0, r_1, \cdots, r_l\}$ denotes the set of all different kinds of user behaviors. According to the number of types of different user behaviors, we construct $|\mathbf{R}|=l+1$ hyper meta-graphs for each user. For the $t$-th hyper meta-graph of user $u_j$, it is defined as $\mathbf{HG}_t^j=\{(u_j, (r_a, r_b, \cdots, r_c), i_q)|i_q\in\mathbf{I}, \forall r\in\{r_0, r_1, \cdots, r_{t-2}, r_l\}\}$, where behavior sequence $(r_a, r_b, \cdots, r_c)$ is sorted in chronological order, and each behavior $r$ in the sequence solely bridges user $u_j$ and item $i_q$. Hence, we will have a set of hyper meta-graphs $\mathcal{HG}^j=\{\mathbf{HG}_0^j, \mathbf{HG}_1^j, \cdots, \mathbf{HG}_l^j\}$ to illustrate user-item interactions of user $u_j$ in the recommendation system. Note that the order among different behaviors in $\mathbf{R}$ is based on the distance between behaviors and behavior \textit{buy} in the semantic space. For example, four types of common user behaviors, \textit{page view}, \textit{favorite}, \textit{add to cart}, \textit{buy}. Behavior \textit{page view} is farthest from behavior \textit{buy}, we can define a sorted set of behaviors here, $\{r_{pv}, r_{fav}, r_{cart}, r_{buy}\}$. So, the first hyper meta-graph of a user solely contains the behavior of \textit{buy}, the second one contains \textit{page view} and \textit{buy}, the third one contains all the behaviors except \textit{add to cart}, and the last hyper meta-graph of the user contains all of four types of behaviors.


\subsection{Graph Encoders}
Graph encoder is the essential part of the whole framework since it determines whether the framework can learn representative embeddings for users' behavior patterns from hyper meta-graphs. GNN models are widely used graph encoders, e.g., GCN \cite{gcn} and GAT \cite{gat}. Technically, any GNN models can be used in our framework with sufficient information (e.g., edge types and node features). Note that we do not need exquisite GNN models since we have built hyper meta-graphs which carry rich semantic information (e.g., geometric information, topological structures). In the practice of our framework, we could apply geometric or topoloy based GNNs, like GIN \cite{gin} and TAG \cite{tag}, as the graph encoder because of their simplicity and effectiveness. Further more, they can leverage the structure information from proposed hyper meta-path and advantages of structure-level contrastive learning \cite{gcc}. 

As mentioned in the previous section, each user in the recommendation system have $|\mathbf{R}|$ hyper meta-graphs. We assign $|\mathbf{R}|$ independent graph encoders to process these hyper meta-graphs accordingly, note that these graph encoders are shared among different users. Given the $t$-th hyper meta-graph of user $u_j$ and a graph encoder $g_t(\cdot)$, where $\cdot$ denotes a hyper meta-graph, we will have the embedding of the $t$-th hyper meta-graph of user $u_j$:
\begin{equation}
	\mathbf{h}^j_t = g_t(\mathbf{HG}^j_t),
\end{equation}
where $\mathbf{h}^j_t \in \mathbf{R}^h$ and $h$ denotes the hidden dimension of the user behavior pattern embeddings and item embeddings.

\subsection{Hyper Meta-Graph Contrastive Learning}
For each user, we build several hyper meta-graphs. The graphs carry the interaction records of the user. We can capture this information via graph encoders learning on the hyper meta-graphs separately. However, the behavior patterns of a user would be complicated. According to the example mentioned in the previous section, we have four different hyper meta-graphs for each user. The complexity of the hyper meta-graph is increasing following the number of behavior types it contains. For example, the first hyper meta-graph solely includes \textit{buy}, and the second hyper meta-graph includes \textit{page view} and \textit{buy}. The second hyper meta-graph contains at least two purchasing patterns: buying the item directly, which is also contained in the first hyper meta-graph, and buying the item after viewing. Suppose we adopt graph encoders to learn on each hyper meta-graph separately. In that case, different behavior patterns in the same hyper meta-graph will be fused. This result may neglect the performances when using the learned behavior patterns for the recommendation. It is critical to extract different behavior patterns from a sequence of hyper meta-graphs whose complexities are cascadingly increasing. A potential solution is to contrast the hyper meta-graph with its previous one to obtain the differences (e.g., different behavior patterns) between these two adjacent hyper meta-graphs.

Thanks to the recent success of contrastive learning in graph learning, we propose to utilize \textit{hyper meta-graphs discrimination} as the solution to obtain different behavior patterns and InfoNCE as the contrastive learning objective.

We give an example here. For the user $u_j$, we give out two adjacent hyper meta-graphs of $u_j$, which are $HG_{t-1}^j$ and $HG_t^j$. We assign $g_{t-1}(\cdot)$ and $g_t(\cdot)$ as their graph encoders, respectively. Hence, we will have embeddings of two hyper meta-graphs:
\begin{equation}
	\mathbf{h}_{t-1}^j=g_{t-1}(\mathbf{HG}^j_{t-1}),
\end{equation}
\begin{equation}
	\mathbf{h}^j_t = g_t(\mathbf{HG}^j_t).
\end{equation}
In this example, $\mathbf{h}_{t-1}^j$ and $\mathbf{h}^j_t$ compose the negative pair. To satisfy the setting of InfoNCE, we must construct the positive pair to fulfill the contrastive learning process. Following the graph contrastive learning settings in GCC \cite{gcc}, we use $g_{t-1}(\cdot)$ to encode $HG_t^j$ to obtain $\mathbf{\hat{h}}_t^j$:
\begin{equation}
	\mathbf{\hat{h}}_t^j=g_{t-1}(\mathbf{HG}^j_{t}),
\end{equation}
which is together with $\mathbf{h}^j_t$ to compose the positive pair. In our work, we adopt InfoNCE such that:
\begin{equation}
	\mathcal{L}^j_{t-1, t} = -\ln\frac{\exp{(d(\mathbf{h}^j_t, \mathbf{\hat{h}}_t^j))}}{\exp{(d(\mathbf{h}^j_t, \mathbf{\hat{h}}_t^j))} + \exp{(d(\mathbf{h}^j_t, \mathbf{h}_{t-1}^j)})},
\end{equation}
where $d(\cdot, \cdot)$ denotes the metrics measuring the distance between two vectors. For the recommendation system having $n+1$ users and $l+1$ different types of user behaviors, we will have a total contrastive learning objective:
\begin{equation}
	\mathcal{L}_{contra} = \frac{1}{n+1} \sum_{j=0}^n \sum_{t=1}^l \mathcal{L}^j_{t-1, t}.
\end{equation}
The intuitions of adopting such a strategy are twofold:
\begin{itemize}
	\item \textbf{Avoid generating the negative pair via graph augmentation}. Some works \cite{graph-aug} utilize graph augmentation to generate negative pairs. But in the recommendation scenario, graph augmentation would disturb the users' interaction records and affect behavior pattern generation, which may cause misleading results in the downstream recommendation tasks. Such a strategy is an alternative solution for us to generate the negative pair without disturbing original semantics.
	\item \textbf{Bridge two contrasting hyper meta-graphs}.  It is hard for us to link the embeddings generated from different graph encoders with different graphs in semantic space. However, with such a strategy, we can build an implicit connection between contrasting hyper meta-graphs in the contrastive learning process.
\end{itemize}
We will have $|\mathbf{R}|$ user behavior embeddings for a user after contrastive learning process, which will be fed into fusion layer and downstream recommendation tasks.

\subsection{Users' Multi-behavior Patterns Fusion}
\label{sec:fusion}
After obtaining $|\mathbf{R}|$ different embeddings which denote different behavior patterns of a user, we have to fuse them and obtain a unified embedding to conduct recommendations. There is a sort of widely used linear fusion methods, like \textit{sum} and \textit{mean}. And there is another type of fusion method, which is neural network-based methods (e.g., Multi-Layer Perceptron (MLP) and Personalized Non-Linear Fusion (PNLF) \cite{personalized-fusion}). Given a fusion function $f(*)$, we can have a unified behavior pattern embedding for the user:
\begin{equation}
	\mathbf{h}^j_{uni} = f(\mathbf{h}^j_0, \mathbf{h}^j_1, \cdots, \mathbf{h}^j_l) \in \mathbf{R}^h.
\end{equation}

\subsection{Recommendation Task}
There are plenty of collaborative filtering-based recommendation frameworks, which leverage the explicit or implicit feedback of users, like \cite{ncf}. To fully demonstrate the ability of the proposed model, we use vector product to make prediction instead of those complex and state-of-the-art models to avoid the improvement brought by the recommendation model.

Let $\mathbf{h}^k$ denotes the embedding of item $i_k$. With the unified behavior pattern embedding of user $u_j$, we can obtain the predicted score between the item and the user via:
\begin{equation}
	\hat{p}_{u_j, i_k}={\mathbf{h}_{uni}^j}^T \boldsymbol{W} \mathbf{h}^k,
\end{equation}
where the trainable weight matrix $\boldsymbol{W} \in \mathbf{R}^h$. The matrix $\boldsymbol{W}$ is used to map the unified behavior pattern to the space where item embeddings in for score prediction.

To train model parameters, we take the negative logarithm \cite{ncf} of the likelihood function:
\begin{equation}
	\begin{aligned}
		\mathcal{L}_{rec} = &-\sum_{(u_j, i_k)\in \mathcal{Y}\cup\mathcal{Y}^-}p_{u_j, i_k}\log\hat{p}_{u_j, i_k}\\
		&+(1-p_{u_j, i_k})\log(1-\hat{p}_{u_j, i_k}),
	\end{aligned}
\end{equation}
to normalize the loss value of loss function on recommendation tasks, we take
\begin{equation}
	\mathcal{L}_{ave\_rec} = \frac{\mathcal{L}_{rec}}{|\{(u_j, i_k)|(u_j,i_k)\in\mathcal{Y}\cup\mathcal{Y}^-\}|}
\end{equation}
as the objective, where $\mathcal{Y}$ and $\mathcal{Y}^-$ denote postive interaction records and sampled negative interaction records, $p_{u_j, i_k}\in\{0, 1\}$ represents if there is an interaction between user $u_j$ and item $i_k$.

To train the model from end to end, we couple two objectives as the total loss function:
\begin{equation}
	\mathcal{L} = (1-\beta)\cdot\mathcal{L}_{contra}+\beta\cdot\mathcal{L}_{ave\_rec},
\end{equation}
where $\beta$ is a hyperparameter contoling the significance of two objectives in the total training objective.

\section{Experiments}
This section evaluates HMG-CR on recommendation tasks with two real-world datasets. We will first report the comparison experiment results of HMG-CR and baselines. Then, we analyze how the graph contrastive learning works in our model. Lastly, we conduct ablation studies on graph encoder and fusion layer in the model.

\subsection{Datasets}
We evaluate the proposed framework on two real-world datasets, which have high quality and are widely used. including Taobao\footnote{https://tianchi.aliyun.com/dataset/dataDetail?dataId=649} and Tmall\footnote{https://tianchi.aliyun.com/dataset/dataDetail?dataId=47}.
To ensure the quality of the datasets, we follow the customary practice \cite{bert4rec} to discard users and items with less than five interactions of \textit{buy}. We also filter users with too much interactions of \textit{page view} in Tmall to discard noise. The statistics of the filtered datasets are shown in Table \ref{tab:statistics}.

\begin{table}
	\renewcommand\arraystretch{0.8}
	\centering
	\caption{Statistics of Datasets}
	\label{tab:statistics}
		\resizebox{0.4\textwidth}{!}{
	\begin{tabular}{ccc}
		\toprule
		Dataset             & Taobao            & Tmall             \\ \midrule
		\#users             & 48946             & 9368              \\
		\#items             & 1500839           & 302722            \\
		\#pv (percentage)   & 7723217 (85.17\%) & 1510303 (92.14\%) \\
		\#fav (percentage)  & 436715 (4.82\%)   & 102419 (6.25\%)   \\
		\#cart (percentage) & 527221 (5.81\%)   & 24557 (1.50\%)    \\
		\#buy (percentage)  & 380877 (4.20\%)   & 104360 (6.37\%)   \\
		\#total             & 9068030           & 1639220           \\ \midrule
		\#ave\_pv           & 157.79            & 161.22            \\
		\#ave\_fav          & 8.92              & 10.93             \\
		\#ave\_cart         & 10.77             & 2.62              \\
		\#ave\_buy          & 7.78              & 11.14             \\
		\#ave\_total        & 185.27            & 174.98            \\ \bottomrule
	\end{tabular}}
\end{table}

\subsection{Baselines}
To verify the effectiveness of the proposed framework, we compare it with three categories of baselines. The first category is conventional GNNs including GCN \cite{gcn} and GraphSAGE \cite{graphsage}, which cannot distinguish different types of edges in the graph, they treat different user behaviors as the same. The second category is edge types-aware GNNs including GAT \cite{gat} and RGCN \cite{rgcn}, which can process various types of edges in the graph explicitly or implicitly to capture the features  of different user behaviors. The last category is novel multi-behavior recommendation frameworks, NMTR \cite{nmtr} and EHCF \cite{ehcf} which achieve state-of-the-art performances on multi-behavior recommendation tasks.

\subsection{Experiment Settings}
For reproducibility,  we introduce the details of the hyperparameter settings of the proposed framework. We train our model on dataset Taobao with learning rate $lr=0.0001$, weight decay $wd=0.000001$, hidden dimension $h=16$ advised by \cite{where}, and 3-layer TAG as graph encoder. As to dataset Tmall, we tune our model with the same learning rate, weight decay and hidden dimension. We take 3-layer GIN as graph encoder for dataset Tmall. To ensure the fairness in the comparison studies, we follow the widely used \textit{leave-one-out} strategy \cite{ncf} to conduct comparison. 
The metrics we adopt are Recall@K and NDCG@K, which show the recommendation quality of top-K recommended items. For more details, you can refer to the source code of this project via this link\footnote{https://github.com/Haoran-Young/HMG-CR}. All experiments are conducted on NVIDIA TITAN Xp.

\subsection{Comparison Experiment Results}
\begin{table*}[htbp]
	\footnotesize
	\renewcommand\arraystretch{0.4}
	\centering
	\caption{Comparison experiment results of HMG-CR and baselines}
	\label{tab:results}
	\resizebox{0.8\textwidth}{!}{
		\begin{tabular}{c|cccc|cccc}
			\toprule
			Dataset            & \multicolumn{4}{c|}{Taobao}                                                                                                                     & \multicolumn{4}{c}{Tmall}                                                                                                                       \\ \midrule
			\diagbox{Methods}{Metrics} & Recall@5 & Recall@10 & NDCG@5 & NDCG@10 & Recall@5 & Recall@10 & NDCG@5 & NDCG@10 \\ \hline
			GCN                & 0.2577                             & 0.3589                    & 0.1842                           & 0.2167                            & 0.2544                             & 0.3775                              & 0.1763                           & 0.2163                            \\
			GraphSAGE          & 0.2751                             & 0.3826                              & 0.1965                           & 0.2312                            & 0.2588                             & 0.3695                              & {\ul 0.1813}                     & 0.2170                            \\
			GAT                & 0.2782                             & 0.3921                              & {\ul 0.1972}                     & 0.2339                            & 0.2561                    & 0.3735                    & 0.1777                           & 0.2158                            \\
			RGCN               & 0.2714                             & 0.3767                              & 0.1946                           & 0.2285                   & 0.2725                             & 0.4144                              & 0.1749                           & 0.2215                            \\
			NMTR               & 0.2215                   & 0.3781                              & 0.1513                           & 0.2012                            & {\ul 0.2780}                       & {\ul 0.4230}                     & 0.1798                           & {\ul 0.2265}                      \\
			EHCF               & {\ul 0.2882}                       & {\ul 0.4166}             & 0.1945                  & {\ul 0.2359}                      & 0.2451                             & 0.4115                              & 0.1581                           & 0.2113                            \\ \midrule
			HMG-CR(SG)         & 0.3050                             & 0.4417                              & 0.2162                           & 0.2608                            & 0.2943                             & 0.4329                              & 0.1863                  & 0.2321                  \\
			HMG-CR(GCN)        & 0.3039                             & 0.4441                              & 0.2154                           & 0.2613                            & 0.2954                             & 0.4332                              & 0.1869                           & 0.2324                            \\ \midrule
			HMG-CR(GAT)        & 0.3460                             & 0.4390                              & 0.2443                           & 0.2746                            & 0.3163                             & 0.4320                              & 0.2224                           & 0.2604                            \\ \midrule
			HMG-CR(GIN)        & 0.3141                             & 0.3627                              & 0.2029                           & 0.2191                            & \textbf{0.3547}                    & 0.4313                              & \textbf{0.2642}                  & \textbf{0.2891}                   \\
			HMG-CR(TAG)        & \textbf{0.3588}                    & \textbf{0.4464}                     & \textbf{0.2639}                  & \textbf{0.2926}                   & 0.2964                             & \textbf{0.4350}                     & 0.1902                           & 0.2359                            \\ \midrule
			Improvement        & 24.50\%                            & 7.15\%                              & 33.82\%                          & 24.04\%                           & 27.59\%                            & 2.84\%                              & 45.73\%                          & 27.64\%                          \\ \bottomrule
		\end{tabular}
	}
\end{table*}

Table \ref{tab:results} lists the comparison experiment results for all methods on two datasets. Overall, the proposed  framework HMG-CR with different graph encoders consistently and significantly outperforms all baselines in terms of all metrics. Particularly, our proposed framework has more significant improvement on the metric NDCG, which shows that our proposed framework pays more attention to sorting recommended items. Note that HMG-CR on dataset Taobao slightly outperforms that on dataset Tmall. According to the statistics of the two datasets, as shown in Table \ref{tab:statistics}, we note that the average numbers of total interactions for each user are close in two datasets, but there are differences among the distribution of numbers of different user behaviors. The ratio of \textit{add to cart} in dataset Tmall is much less than that in dataset Taobao. Each user in both datasets has four hyper meta-graphs since there are four different types of user behaviors. Due to lack of \textit{add to cart} in dataset Tmall, the third hyper meta-graph for a user, including \textit{page view}, \textit{add to cart}, and \textit{buy}, is similar to the second hyper meta-graph for the user, including \textit{page view} and \textit{buy}. Under such a scenario, it is hard for graph contrastive learning to maximize the dissimilarities between the second hyper meta-graph and the third hyper meta-graph. Hence, the user behavior pattern embedding generated in this part would be misleading for unified user behavior pattern embedding generation.

Graph neural network-based methods have unsatisfying performances in our experiment. The interaction graphs for each user in the recommendation systems have simple structures (e.g., bipartite graphs). Conventional GNN models, like GCN and GraphSAGE, may be insufficient to capture user behavior pattern embeddings on such simple graph structures. Edge types-aware GNN models, like GAT and RGCN, slightly outperform GCN and GraphSAGE since they integrate fruitful side information regarding different types of user behaviors. Overall, two categories of GNN models have no significant gaps, because \textit{page view} takes the most of place in the datasets. Message passing and aggregation are not capable to capture sophisticated relations among different types of user behaviors, since semantics of \textit{page view} would conceal other information.

NMTR and EHCF are state-of-the-art multi-behavior recommendation frameworks. They leverage the well-designed recommendation models and multi-task learning strategy to utilize the supervision signals from all types of user behaviors. However, there is a limitation for both frameworks. Both of them have an assumption that each type of user behaviors has strong connections with precedent types of user behaviors. This assumption is not solid, because users' behavioral patterns are complex as shown in a toy example in Section \ref{sec:hmp}. The proposed HMG-CR adopts a more flexible manner to utilize graph contrastive learning to capture the dependencies among different types of user behaviors instead of assuming there are strong  connections between a behavior and the precedent one. Because of it, even without multi-task supervision signals and well-designed recommendation models, the proposed HMG-CR still outperforms NMTR and EHCF with leveraging the advantages of hyper meta-graphs and graph contrastive learning.

\subsection{Analysis of Graph Contrastive Learning}
\begin{figure}[htbp]
	\centering
	\includegraphics[width=0.25\textwidth]{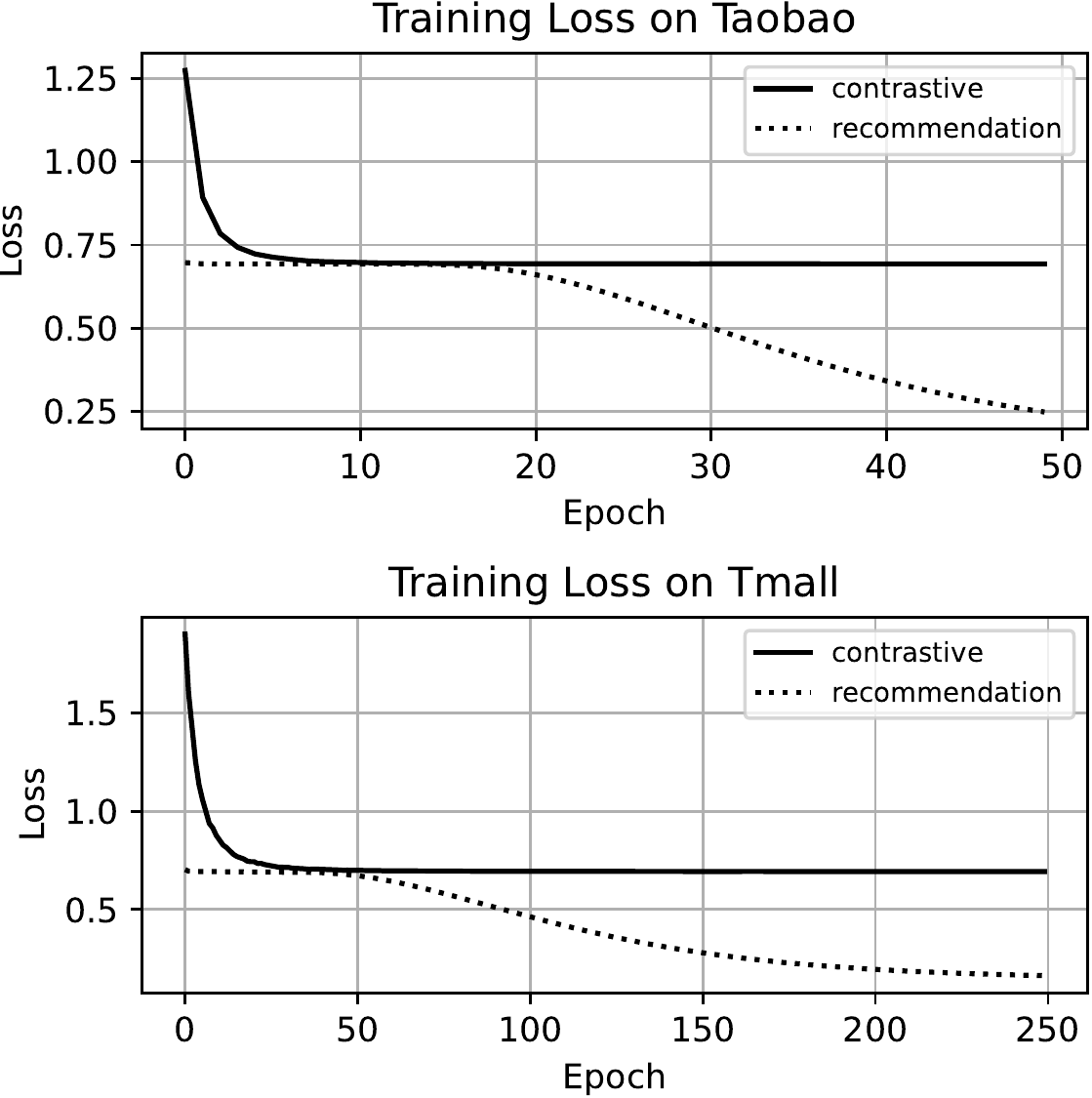}
	\caption{The contrastive loss and recommendation loss of HMG-CR on both datasets in training process}
	\label{training_loss}
\end{figure} 

In this section, we will introduce to you the detailed mechanism of graph contrastive learning in our proposed framework. First, as shown in the Figure \ref{training_loss}, we demonstrate the training loss of the proposed framework on two datasets during the training process. The training loss is twofold, contrastive loss and recommendation loss. We observe a clear tendency that contrastive loss drops first and remains stabilized, and then the recommendation loss starts to decrease. This phenomenon reflects that our proposed framework first maximizes the dissimilarity among hyper meta-graphs to obtain user behavior pattern embeddings and updates parameters on the recommendation task. And the contrast among hyper meta-graphs is maintained within the whole training process. 

\begin{figure}[htbp]
	\centering
	\subfigure[]{
		\includegraphics[width=0.2\textwidth]{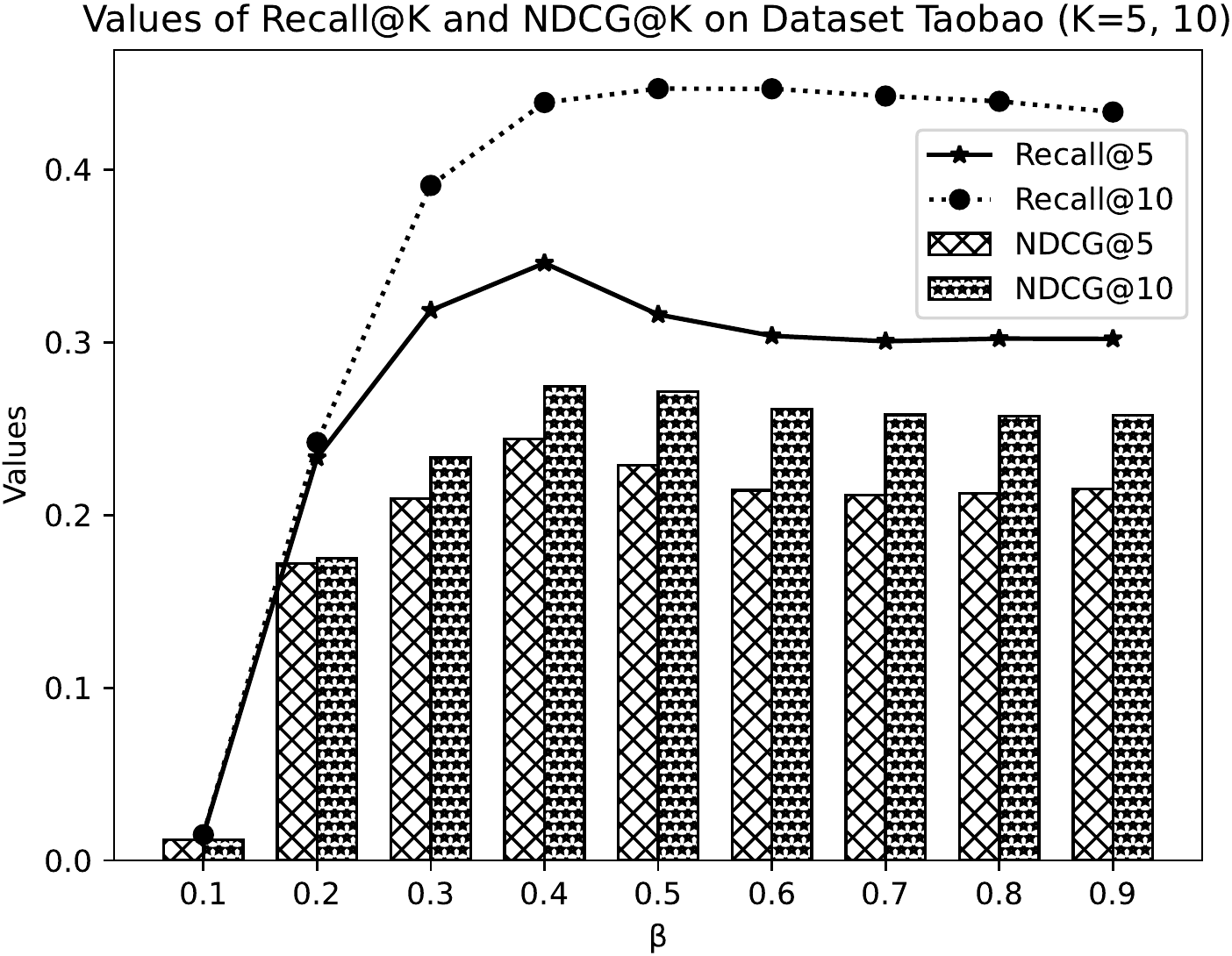}
	}
	\quad
	\subfigure[]{
		\includegraphics[width=0.2\textwidth]{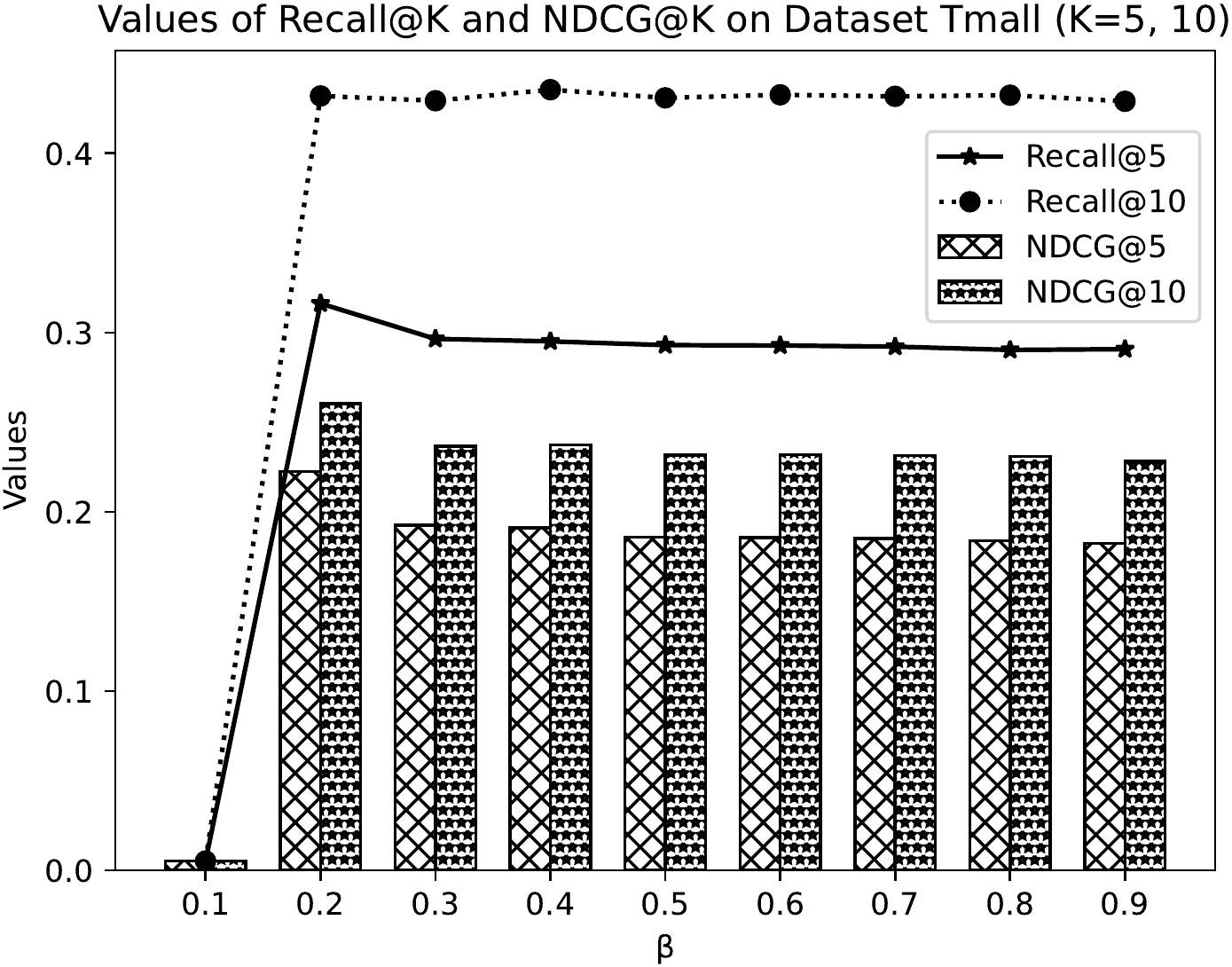}
	}
	\caption{(a) Values of Recall@K and NDCG@K on dataset Taobao variation trend with different $\beta$. (b) Values of Recall@K and NDCG@K on dataset Tmall variation trend with different $\beta$. (Graph encoder and fusion layer of HMG-CR here are GAT and MEAN, respectively.)}
	\label{fig:beta}
\end{figure}

To further illustrate the impact of graph contrastive learning on our proposed framework, we conduct hyperparameter studies on $\beta$, which controls the relative significance of graph contrastive learning tasks and recommendation tasks. The experimental results of hyperparameter studies are shown in the Figure \ref{fig:beta}. Overall, our proposed framework is not that sensitive to $\beta$ as long as $\beta$ is not too small. However, we note that the proposed framework has worse results when $\beta$ takes the boundary values (e.g., $\beta=0.1$). When $\beta$ is too low, the model pays less attention to the recommendation tasks. The model cannot acquire sufficient supervision signals from training data to update the parameters. Under this scenario, it is difficult for our model to converge quickly and precisely on recommendation tasks. With $\beta$ increasing, the performances of the proposed framework increase accordingly. When $\beta$ is larger than some specific values, e.g., $\beta=0.4$ for dataset Taobao and $\beta=0.2$ for dataset Tmall, the performances start to decrease slightly. Large $\beta$ values will neglect graph contrastive learning tasks, which would undermine the ability of the model to acquire user behavior pattern embeddings from sophisticated hyper meta-graphs. This phenomenon also verifies that graph contrastive learning is helpful to our proposed framework. In summary, graph contrastive learning tasks and recommendation tasks should have a relatively balanced significance and $\beta$ should not be too large in our proposed framework to avoid decreasing model performances and cannot be too small in which case the framework may not work.

\subsection{Ablation Studies}
\label{sec:flexibility}

\begin{table*}[]
	\footnotesize
	\renewcommand\arraystretch{0.4}
	\centering
	\caption{Supplementary experiment results of GIN and TAG}
	\label{tab:supplementary}
	\resizebox{0.8\textwidth}{!}{
		\begin{tabular}{c|cccc|cccc}
			\toprule
			Dataset            & \multicolumn{4}{c|}{Taobao}                                                                                                                     & \multicolumn{4}{c}{Tmall}                                                                                                                       \\ \midrule
			\diagbox{Methods}{Metrics} & Recall@5& Recall@10 & NDCG@5 & NDCG@10 & Recall@5 & Recall@10 & NDCG@5 & NDCG@10 \\ \midrule
			GIN                & 0.2682                             & 0.3779                              & 0.1892                           & 0.2246                            & 0.2817                             & 0.4236                              & 0.1878                           & 0.2340                            \\
			TAG                & 0.2784                             & 0.3863                              & 0.1994                           & 0.2342                            & 0.2845                             & 0.4235                              & 0.1869                           & 0.2323                            \\ \midrule
			HMG-CR(GIN)        & 0.3141                             & 0.3627                              & 0.2029                           & 0.2191                            & \textbf{0.3547}                    & 0.4313                              & \textbf{0.2642}                  & \textbf{0.2891}                   \\
			HMG-CR(TAG)        & \textbf{0.3588}                    & \textbf{0.4464}                     & \textbf{0.2639}                  & \textbf{0.2926}                   & 0.2964                             & \textbf{0.4350}                     & 0.1902                           & 0.2359                            \\ \bottomrule
		\end{tabular}
	}
\end{table*}

\begin{table*}[htbp]
	\footnotesize
	\renewcommand\arraystretch{0.4}
	\centering
	\caption{Performances of HMG-CR(GAT) with different fusion layers on both datasets\\ ($\beta=0.4$ for dataset Taobao and $\beta=0.2$ for dataset Tmall)}
	\label{tab:fusion}
	\resizebox{0.8\textwidth}{!}{
		\begin{tabular}{c|cccc|cccc}
			\toprule
			Dataset           & \multicolumn{4}{c|}{Taobao}                                                                                             & \multicolumn{4}{c}{Tmall}                                                                                               \\ \midrule
			\diagbox{Fusion}{Metrics} & Recall@5 & Recall@10 & NDCG@5 & NDCG@10 & Recall@5 & Recall@10 & NDCG@5 & NDCG@10 \\ \hline
			MEAN              & \textbf{0.3460}              & 0.4390                        & \textbf{0.2443}            & \textbf{0.2746}             & \textbf{0.3163}              & 0.4320                        & \textbf{0.2224}            & \textbf{0.2604}             \\
			SUM               & 0.3012                       & \textbf{0.4427}               & 0.2118                     & 0.2580                      & 0.2939                       & 0.4345                        & 0.1879                     & 0.2343                      \\ \midrule
			MLP               & 0.3024                       & 0.4344                        & 0.2150                     & 0.2579                      & 0.2946                       & \textbf{0.4349}               & 0.1873                     & 0.2336                      \\
			PNLF              & 0.3046                       & 0.4363                        & 0.2157                     & 0.2586                      & 0.2944                       & 0.4344                        & 0.1865                     & 0.2327                      \\ \bottomrule
		\end{tabular}
	}
\end{table*}

\noindent\textbf{Graph Encoder} Choosing a proper graph encoder for the framework determines whether it can achieve good performances. We select three common categories GNNs, conventional message passing based GNNs including SG \cite{sg} and GCN, attention mechanism based GNNs including GAT, and graph topological or geometric structure aware GNNs including TAG \cite{tag} and GIN \cite{gin}. The experiment results are shown in the Table \ref{tab:results}. According to the results, the proposed HMG-CR with any graph encoders outperforms all baselines. Specifically, HMG-CR with SG or GCN slightly outperforms baselines since conventional message passing based GNNs are insufficient to capture complex user behavior features from the constructed hyper meta-graphs. Despite the user-item interactions in the hyper meta-graphs, there are also chronological dependencies among different user behaviors. With such sophisticated relations in the hyper meta-graphs, GAT leverages attention mechanism to learn user behavior embeddings via adaptively distinguish different relations (edges) in the hyper-meta graphs. However, we replaced different types of edges, which represent user behaviors, with different types of nodes in the hyper meta-graphs. We explicitly add the information of interactions among users and items into the hyper meta-graph. It means that the improvement brought by attention mechanism, distinguishing different edges is limited. Note that the hyper meta-graphs have a structure which is similar to tree topology. Hence, the hyper meta-graphs have not only fruitful semantic information but also excellent structure. HMG-CR with graph structure aware graph encoders leverages the advantages of the hyper meta-graphs and achieve the best results in our experiments. To verify the improvement is brought by our proposed framework instead of TAG or GIN solely, we conduct supplementary experiments with TAG and GIN shown in the Table \ref{tab:supplementary}. 

\noindent\textbf{Fusion Layer} Fusion layer is the output layer of the proposed framework. We focus on two categories of fusion layers, linear fusion layer, \textit{mean} and \textit{sum}, and non-linear fusion layer, MLP and PNLF. The experimental results are shown in the Table \ref{tab:fusion}. According to the results, HMG-CR taking \textit{mean} as the fusion layer achieves the best result. 
Overall, HMG-CR with linear fusion layer has better performances in our experiments. We note that there are mapping layers in MLP and PNLF. 
In this component, mapping mechanism may disturb the  user behavior pattern obtained in the space in which the graph contrastive learning was conducted. Hence, we should take linear fusion layer to output the unified user behavior pattern embeddings to avoid disturbing caused by conducting fusion in another embedding space.

\section{Related Works}
\label{sec:relatedworks}
\subsection{Graph Contrastive Learning}
Graph contrastive learning recently attracts attention from researchers leveraging contrastive learning idea to enhance existing GNNs. The core idea of contrastive learning is to maximize representation agreement among sampled and transformed data. Some works \cite{gcc, graph-aug} introduce contrastive learning to graph representation learning and had achieved promising results. There are two main measures to generate contrasting pairs, including graph perturbation and sub-graph sampling. However, how to adpatively construct contrasting pairs instead of randomly and how to implement graph contrastive learning in real-world problems are fully explored. 

\subsection{Multi-behavior Recommendation}
Multi-behavior recommendation utilizes multiple user-item feedback for enhancing recommendation on target behaviors. There are different approaches to make use of users' multi-behavior information. In a multi-behavior interactions graph, \cite{gat, rgcn, kg-rec} assign different weights to different types of edges, representing different types of behaviors, before conducting aggregation. Graph based recommendation methods \cite{gat, rgcn, hyper-collab-rec} achieve good performances in recommendation tasks with leveraging advantages of GNNs. Moreover, \cite{nmtr, ehcf} adopt multi-task learning techniques to acquire more supervision signals from multi-behavior data. For embedding generation, they assume that one behavior is strongly related to the precedent behaviors and embeddings of different types of user behaviors are adjacent in the embedding space. Both aggregation in \cite{gat, rgcn} and assumptions in \cite{nmtr, ehcf} are insufficient to capture complex relationships among different behaviors of users.

\section{Conclusion}
In this paper, we propose the concept of hyper meta-path and a novel framework, HMG-CR, which first utilizes graph contrastive learning techniques into recommendation systems. Leveraging the advantages of hyper meta-path and HMG-CR, we achieve the SOTA performances on the task of purchasing prediction on both datasets. We also conduct extensive analysis on HMG-CR and fully demonstrate the details of it. The concept of hyper meta-path and the framework are flexible and can be used in other heterogeneous graph mining tasks. 

\section*{Acknowledgment}
This work is supported by the Australian Research Council (ARC)
under Grant No. DP200101374 and LP170100891, and NSF under grants III-1763325, III-1909323,  III-2106758, and SaTC-1930941.

\bibliographystyle{IEEEtran}
\bibliography{sample-base}{}

\end{document}